%% file: main.tex
\documentclass[aip,jap,reprint]{revtex4-1}
\draft
\usepackage{graphicx}
\usepackage{dcolumn}
\usepackage{bm}

\usepackage[utf8]{inputenc}
\usepackage[T1]{fontenc}
\usepackage{mathptmx}
\usepackage{etoolbox}
\usepackage{array}
\usepackage{booktabs}
\usepackage{caption}
\usepackage{float}
\usepackage{dblfloatfix}
\usepackage{tabularx}
\usepackage{amsmath}
\usepackage{xcolor}
\usepackage{soul}

\makeatletter
\def\@email#1#2{%
 \endgroup
 \patchcmd{\titleblock@produce}
  {\frontmatter@RRAPformat}
  {\frontmatter@RRAPformat{\produce@RRAP{*#1\href{mailto:#2}{#2}}}\frontmatter@RRAPformat}
  {}{}
}%
\makeatother

\begin{document}
\input{Sections/Title}
\input{Sections/Abstract}

\input{Sections/Introduction}
\input{Sections/Methods}
\input{Sections/Results}
\input{Sections/Conclusion}

\input{References/References}
\end{document}

%% file: Sections/Title.tex
\title{Cavitated Ag Paste for Cost-Effective Solar Cell}

\author{Donald Intal}
\author{Abasifreke Ebong}
\email[Authors to whom correspondence should be addressed: ]{dintal@charlotte.edu}
\affiliation{Department of Electrical and Computer Engineering,UNC Charlotte, Charlotte, NC 28223, USA}

\author{Dana Hankey}
\author{Marshall Tibbetts}
\affiliation{ACI Materials Inc., Goleta, CA 93117, USA}

%% file: Sections/Abstract.tex
\date{\today}
\begin{abstract}
    This paper reports on the investigation of cavitated silver paste produced by cavitation technology as a cost-effective alternative to traditional three-roll milling (TRM). Cavitation, utilizing high-frequency sound waves, enhances metal paste dispersion, reduces oxidation, extends shelf life, and minimizes waste. Passivated Emitter and Rear Cell (PERC) solar cells made with cavitated silver paste achieved a 21$\%$ energy conversion efficiency, slightly lower than the 22$\%$ efficiency of conventional paste. Cavitated paste produced finer gridlines, reducing silver usage and costs but increasing contact resistance, leading to a lower fill factor. Despite this, cavitation technology shows promise for more efficient and cost-effective solar cell production. Further research is needed to optimize efficiency and resistance, highlighting the potential for cavitation technology in commercial applications.
\end{abstract}
\maketitle

%% file: Sections/Introduction.tex
\section{Introduction}

The continuous drive for advancements in solar cell technology has perpetuated a constant search for innovative materials and manufacturing techniques. A key focus of this search is the development of metallic pastes for front gridlines on solar cells, where cost-effectiveness can significantly impact overall performance and market viability. Traditional methods such as three-roll milling (TRM), ball milling (BM), and high-gravity controlled precipitation (HGCP) have been widely adopted in metallic paste production due to their ability to form effective conductive pathways essential for electron collection. However, these methods face challenges such as suboptimal particle size reduction, high energy consumption, and potential contamination \cite{Stefanić2015Phase,Fuerstenau1995Newer,Pang2013Role,Metzger2009Use}. Recent advancements in paste manufacturing technology have focused on overcoming these limitations. For instance, studies by Haider and Hashmi \cite{haider2014health} and Ebong et al. \cite{ebong2013developing} have explored alternative techniques to enhance efficiency and reduce costs. Despite these efforts, there remains a significant gap in achieving both cost-effectiveness and high performance, which our research on cavitated Ag paste aims to address.

\subsection{Overview of Metal Paste Manufacturing Technology}

Three-roll milling (TRM) is notable for its mechanical dispersion capabilities, where a mixture of paste and solvent is processed between three horizontally aligned rolls. The resulting shear forces effectively break down and distribute the particles, making TRM appreciated for its simplicity and the consistent dispersion it achieves\cite{electronic2004electronic,prudenziati1994thick,ha2019effect,nienow1997mixing}. Despite these advantages, TRM faces limitations such as suboptimal particle size reduction and substantial energy demands, challenging its broader applicability and economic efficiency \cite{ebong2013developing,fischer1950colloidal}. 

Ball milling (BM) is another prevalent method. It utilizes metallic or ceramic balls within a rotating drum to achieve particle size reduction through intense collision \cite{haider2014health, zhao2010preparation, gou2012processing, petridis2019graphene}. This technique is renowned for producing pastes with superior printing resolution and electrical conductivity. However, BM's drawbacks include significant energy consumption and the risk of contamination from ball wear, presenting a dilemma between operational efficiency and product purity \cite{sarwat2017contamination, el-eskandarany2015controlling}.

As the metallic paste manufacturing sector progresses, innovative strategies are emerging to overcome the limitations of conventional methods. High-gravity controlled precipitation (HGCP) is a promising alternative, heralding a new era in the synthesis of high-performance silver pastes and offering enhanced manufacturing efficiency \cite{han2016synthesis,ng2012green}. HGCP utilizes a high-gravity environment, typically achieved using a rotating packed bed reactor, to create significant centrifugal forces that enhance mass transfer rates and particle formation. This method allows for the rapid and uniform precipitation of metallic particles from a solution, resulting in fine, uniformly dispersed particles with controlled size and morphology. The increased control over particle characteristics leads to improved paste performance, particularly in terms of conductivity and printability. Additionally, HGCP reduces the need for extensive post-processing, thereby saving energy and reducing costs \cite{bao2019large}. However, HGCP faces challenges such as the need for specialized equipment and maintenance, potential scaling issues, and the complexity of optimizing process parameters for different materials \cite{chen2000synthesis}.

In light of these advancements, cavitation technology marks a groundbreaking approach that leverages high-frequency sound waves to induce the formation and collapse of micro-bubbles within a liquid medium. This process results in highly effective mixing and dispersion of paste constituents, far surpassing traditional methods in both performance and cost-efficiency. The benefits of employing cavitation technology extend well beyond achieving uniform particle distribution; it also significantly curtails oxidation, extends the paste's shelf life by eliminating air bubbles, and reduces waste—collectively leading to substantial cost savings \cite{huneycutt2022cavitated, davidson2021incompressible, blake1996cavitation, baumann2010deaeration}. The following section provides details on the materials and technology involved in cavitation.

\subsection{Cavitation Technology}

Cavitation technology, a fundamental aspect of fluid dynamics, plays a crucial role in the innovative manufacturing of metallic pastes for screen-printed solar cells. It is defined by the formation, growth, and rapid collapse of vapor-filled cavities or bubbles within a liquid medium. The process begins with nucleation, where small gas or vapor pockets emerge as the liquid experiences a pressure drop. These initial cavities are pivotal, setting the stage for subsequent cavitation bubbles \cite{brennen2007phase}. As these bubbles undergo rectified diffusion—expanding and contracting with changes in the liquid's pressure—they absorb surrounding gas, promoting growth \cite{davidson2021incompressible,blake1996cavitation}. Theoretical and numerical studies, such as those conducted by H. Alehossein and Z. Qin \cite{alehossein2007numerical}, provide deep insights into bubble dynamics and their responses to various pressures and conditions. The collapse of these bubbles is particularly significant; it produces extreme temperatures and pressures, emitting shockwaves capable of mechanical disruption and facilitating chemical reactions \cite{blake1996cavitation}.

The Rayleigh-Plesset Equation elegantly models the behavior of cavitation bubbles in a compressible fluid. This equation takes into account several variables, including bubble radius (\(R\)), the rate of change of bubble radius (\(\dot{R}\)), the acceleration of bubble radius (\(\ddot{R}\)), liquid density (\(\rho\)), viscosity (\(\mu\)), surface tension (\(\sigma\)), and surrounding pressures (\(P_g\) for gas pressure inside the bubble, \(P_\infty\) for the far-field pressure, and \(P_v\) for vapor pressure inside the bubble), offering a comprehensive view of the forces at play during cavitation \cite{franc2007rayleigh,plesset1977bubble,tian2008numerical,lathrop1991two,rayleigh2018rayleigh}.

\begin{equation}
    R\ddot{R} + \frac{3}{2} \dot{R}^2 = \frac{1}{\rho} \left( P_g - P_\infty - P_v - \frac{4\mu\dot{R}}{R} - \frac{2\sigma}{R} \right)
\end{equation}

In the production of metallic pastes for solar cells, cavitation principles are ingeniously applied to generate powerful shear forces that break down and evenly distribute metallic particles within the paste, thus enhancing its suitability for solar cell applications. The introduction of ultrasonic waves to the paste creates pressure variations that drive the nucleation, growth, and implosion of cavitation bubbles. The resulting shockwaves achieve a homogeneous distribution of particles, significantly reducing their size and improving the paste's overall properties.

\begin{table}[ht]
\caption{Reduction in Ag Powder Particle Size Due to Cavitation Process}
\label{tab:tab1}
\centering
\begin{tabular}{lD{.}{.}{2}D{.}{.}{2}D{.}{.}{2}}
\toprule
\textbf{Sample} & \multicolumn{1}{c}{\textbf{D10 ($\mu$m)}} & \multicolumn{1}{c}{\textbf{D50 ($\mu$m)}} & \multicolumn{1}{c}{\textbf{D90 ($\mu$m)}} \\
\midrule
Pre-Cavitation & 10.22 & 24.92 & 46.60 \\
Post-Ag - 1    & 0.23  & 0.73  & 1.51  \\
Post-Ag - 2    & 0.20  & 0.65  & 1.43  \\
Post-Ag - 3    & 0.27  & 0.77  & 6.40  \\
Post-Ag - 4    & 0.17  & 0.60  & 1.64  \\
\bottomrule
\end{tabular}
\end{table}

The substantial decrease in particle size, demonstrated in the analysis of Ag powder pre- and post-cavitation, is shown in Table \ref{tab:tab1}. This table clearly demonstrates the technology’s effectiveness in achieving ultra-fine dispersions. As depicted, the D10, D50, and D90 values significantly decrease post-cavitation for all samples. For instance, the D10 value decreases from 10.22 $\mu$m pre-cavitation to as low as 0.17 $\mu$m post-cavitation. Similarly, the D50 value drops from 24.92 $\mu$m to 0.60 $\mu$m, and the D90 value reduces from 46.60 $\mu$m to 1.64 $\mu$m in the most effective cases. These reductions illustrate the technology's capability to achieve finer and more uniform particle distributions, which not only enhances the efficiency of solar cells by improving dispersion quality but also contributes to their cost-effectiveness. The particle sizes in Table \ref{tab:tab1} were measured using Malvern Mastersizer, which utilizes laser diffraction technology to determine the particle size distribution by analyzing the scattering pattern of a laser beam passing through a dispersed particulate sample.

\begin{figure}[ht]
\centering
\includegraphics[width=\columnwidth]{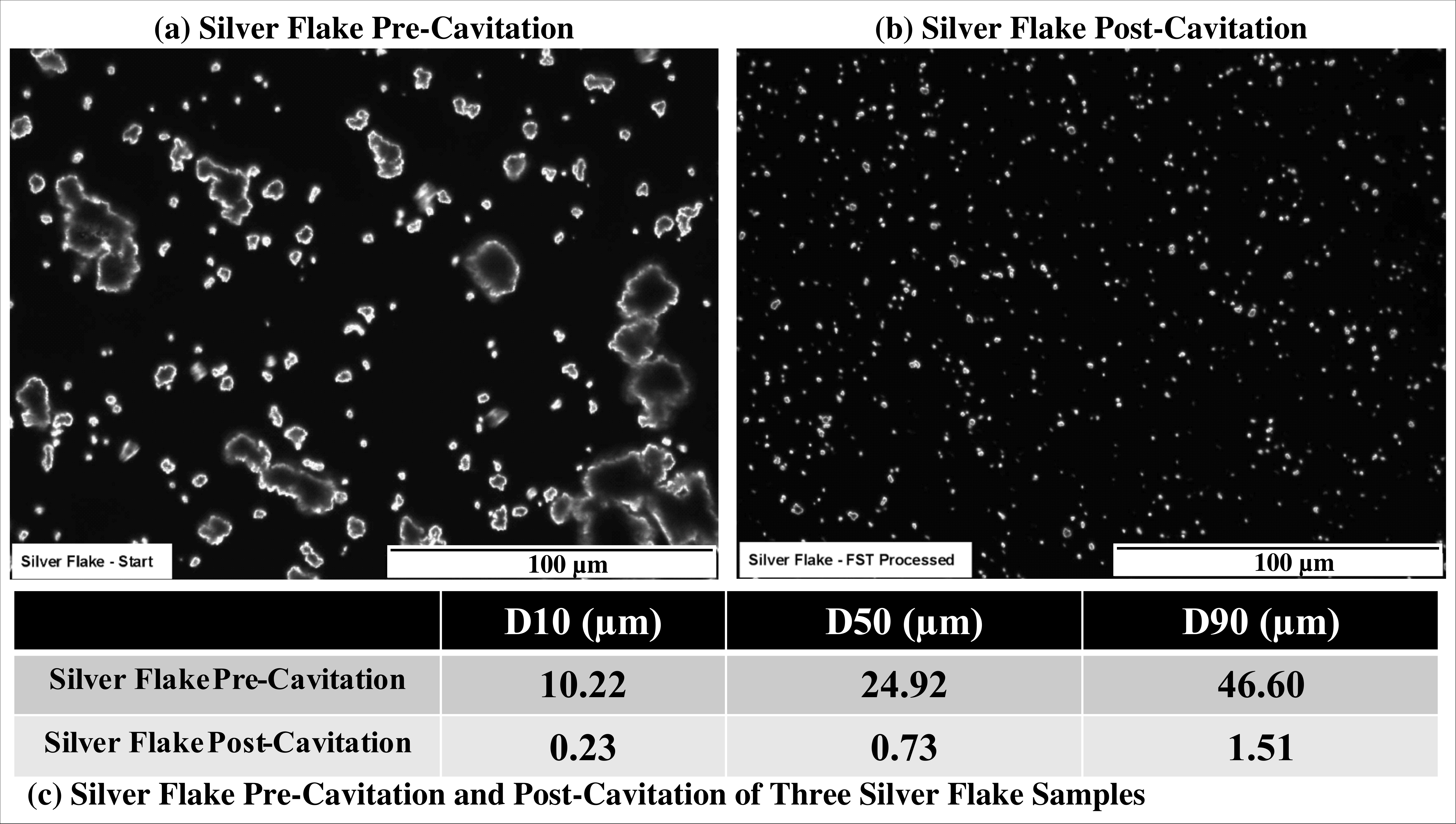}
\caption{Comparative SEM images and particle size distribution of silver flakes before and after cavitation: (a) before cavitation, (b) after cavitation, and (c) particle size distribution table.}
\label{fig:fig1}
\end{figure}

Figure \ref{fig:fig1} presents a comparative SEM analysis, highlighting the dramatic reduction in silver flake particle size after cavitation processing. The images and particle size distribution table clearly demonstrate the substantial effect of cavitation on particle morphology. This significant reduction in particle size, from D10 of 10.22 $\mu$m, D50 of 24.92 $\mu$m, and D90 of 46.60 $\mu$m before cavitation to D10 of 0.23 $\mu$m, D50 of 0.73 $\mu$m, and D90 of 1.51 $\mu$m after cavitation, represents a critical advancement. Such improvements are pivotal in enhancing the efficiency and performance of screen-printed solar cells.

\begin{figure}[ht]
\centering
\includegraphics[width=\columnwidth]{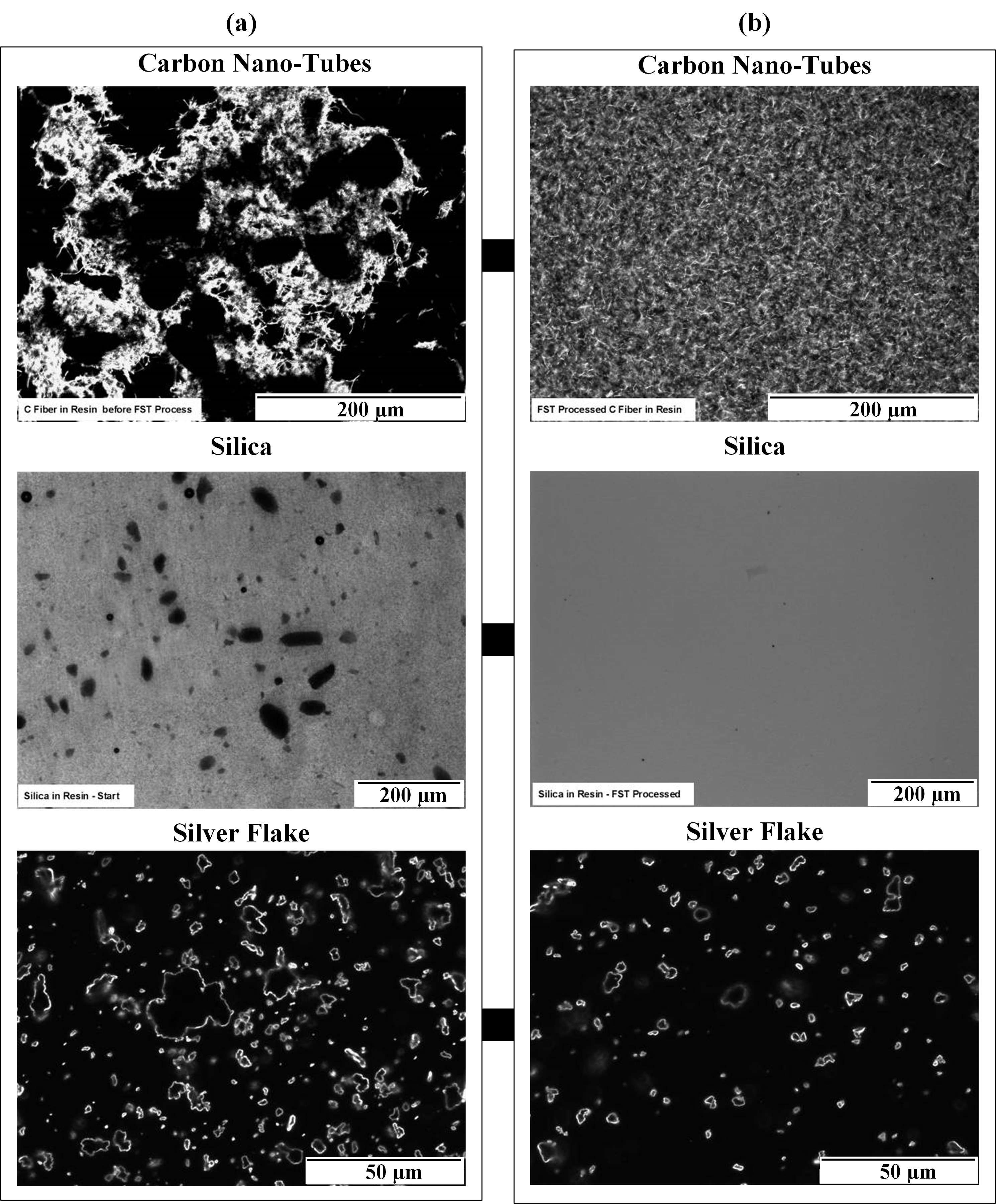}
\caption{Comparative SEM Analysis of Metallic Paste Constituents. (a) shows Silver Flake, Silica, and Carbon Nanotubes before cavitation, (b) displays the same materials post-cavitation.}
\label{fig:fig2}
\end{figure}

Complementing this, Figure \ref{fig:fig2} further details the transformative influence of cavitation technology, highlighting the pronounced uniformity and dispersion of materials. This indicates that while the particles are significantly reduced in size, their form factor is effectively preserved post-cavitation. This dual enhancement in particle size and distribution is pivotal for optimizing the conductive properties of Ag paste.

\begin{figure}[ht]
\centering
\includegraphics[width=\columnwidth]{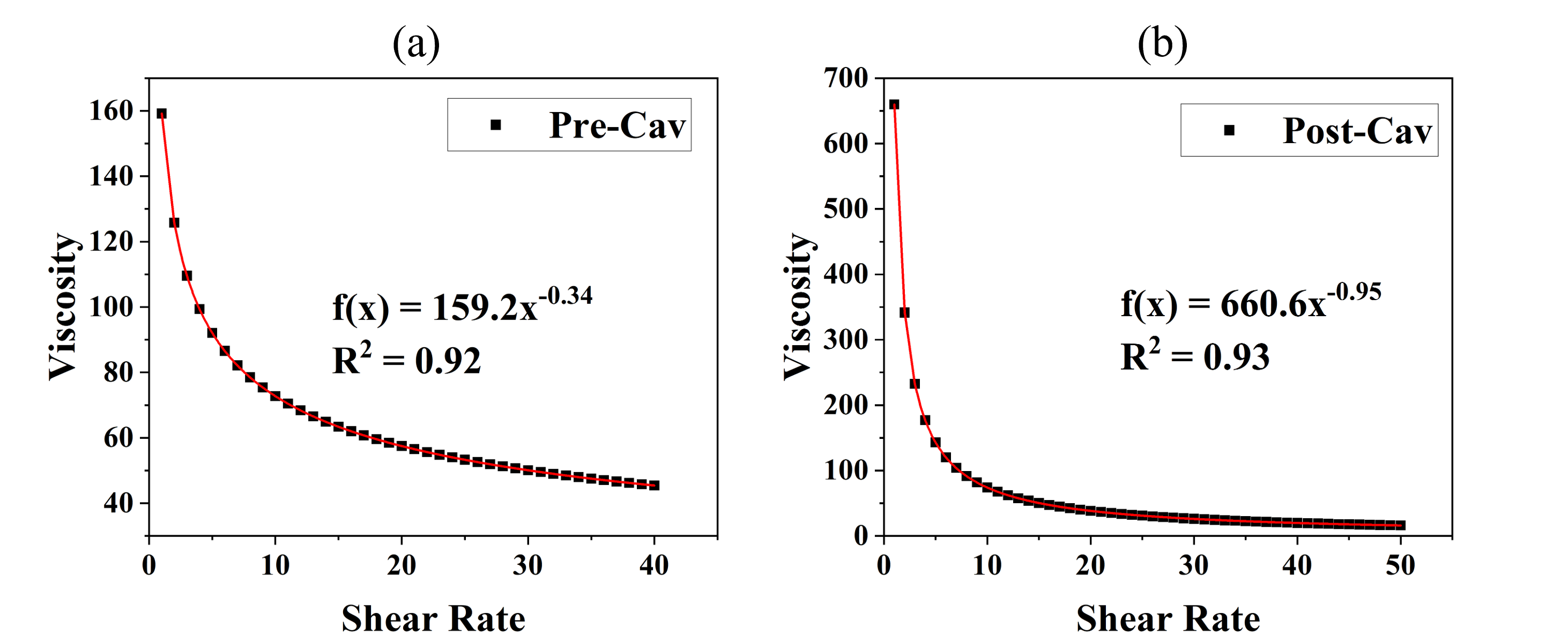}
\caption{Shear rate versus viscosity for Ag pastes. (a) Pre-cavitation (b) Post-cavitation}
\label{fig:fig4}
\end{figure}

Figure \ref{fig:fig4} displays the shear-thinning behavior of Ag pastes, showcasing the differences between pre-cavitation and post-cavitation samples. The pronounced shear-thinning behavior observed in the post-cavitation paste indicates substantial microstructural changes due to cavitation, enhancing the paste's spreadability on substrates and facilitating the creation of finer, more uniform grid lines \cite{yuce2017challenges,schneider2016highly}. The power regression equations ($f(x) = 159.2 x^{-0.34}$ for pre-cav and $f(x) = 660.6 x^{-0.95}$ for post-cav) and their respective $R^2$ values (0.92 and 0.93) indicate how viscosity decreases with increasing shear rate. The higher exponent in the post-cavitation sample’s equation suggests a more pronounced shear-thinning behavior, implying that the cavitation process significantly improves the paste's viscosity response to shear stress, thereby enhancing its application properties. This improvement is beneficial for achieving better printability and more efficient sintering performance.

%% file: Sections/Methods.tex
\section{Cell Fabrication and Characterization}

Commercial (TRM-Ag) and cavitated (Cav-Ag) Ag pastes were used to fabricate Passivated Emitter and Rear Cells (PERC) on G1 blue wafers. Each paste was printed, dried in an infrared belt dryer for two minutes, and sintered in an infrared belt furnace at a peak temperature of 780°C with a belt speed of 533 CPM (210 IPM). Following this, the cells' electrical output was measured using the LIV system from Sinton Instruments, which evaluates current density (J$_{SC}$), voltage (V$_{OC}$), fill factor (FF), and efficiency ($\eta$). Contact resistance was also measured using the Transmission Line Model (TLM) \cite{dobrzanski2010investigation,gregory2019nondestructive,kokbudak2017contact} to quantify its contribution to the total series resistance. This method involves placing metal contacts at various distances on the fingers to measure the resistance directly under the metal contact. Additionally, Electroluminescence (EL) imaging was carried out to qualitatively assess the overall series resistance, detect defects, and check uniformity. This technique is crucial for revealing how uniformly the electrical contacts are formed with the different pastes. 

%% file: Sections/Results.tex
\section{Result and Discussion}

Cavitation technology demonstrated notable improvements in the properties and performance of silver paste used in solar cells compared to the traditional three-roll milling (TRM) method. The open-circuit voltage (V$_{OC}$) of Cav-Ag paste showed a slight improvement, averaging 667 mV, compared to 662 mV for TRM-Ag paste (Table \ref{tab:tab2}). This improvement is attributed to the finer gridlines produced by Cav-Ag, which reduce metal recombination losses. Although the practical impact of the V$_{OC}$ improvement is minimal, the consistency in performance, as indicated by the lower standard deviation (0.001 for Cav-Ag vs. 0.003 for TRM-Ag), suggests potential benefits in large-scale manufacturing, where uniformity is crucial.

\begin{table}[ht!]
\centering
\caption{Comparative Analysis of Cav-Ag and TRM-Ag Metallization Pastes}
\label{tab:tab2}
\begin{tabularx}{\columnwidth}{lXcccccc}
\toprule
\textbf{Variable} & \textbf{Batch\_ID} & \textbf{N} & \textbf{Mean} & \textbf{StDev} & \textbf{Minimum} & \textbf{Maximum} \\
\midrule
Voc (V) & Cav-Ag & 12 & 0.667 & 0.001 & 0.666 & 0.669 \\
 & TRM-Ag & 12 & 0.662 & 0.003 & 0.654 & 0.666 \\
Isc (A) & Cav-Ag & 12 & 10.20 & 0.034 & 10.15 & 10.25 \\
 & TRM-Ag & 12 & 10.14 & 0.056 & 10.03 & 10.22 \\
FF (\%) & Cav-Ag & 12 & 77.42 & 1.544 & 73.80 & 79.07 \\
 & TRM-Ag & 12 & 79.05 & 1.956 & 73.07 & 80.28 \\
Efficiency (\%) & Cav-Ag & 12 & 21.43 & 0.420 & 20.47 & 21.88 \\
 & TRM-Ag & 12 & 21.58 & 0.541 & 20.08 & 22.12 \\
\bottomrule
\end{tabularx}
\end{table}

The short circuit current (I$_{SC}$) for Cav-Ag was slightly higher, averaging 10.20 A compared to 10.14 A for TRM-Ag (Table \ref{tab:tab2}). This increase is likely due to the reduced shading from the finer gridlines. Statistical analysis revealed a significant difference in I$_{SC}$ means (p-value = 0.008), highlighting Cav-Ag's advantage in producing consistent and high-performing cells, with a lower standard deviation (0.034 A for Cav-Ag vs. 0.056 A for TRM-Ag) (Table \ref{tab:tab2}).

The fill factor (FF) of TRM-Ag was higher, averaging 79.05$\%$ compared to 77.42$\%$ for Cav-Ag (Table II). This difference is attributed to the increased contact resistance in Cav-Ag. However, Cav-Ag exhibited more consistent performance with a narrower range (73.80$\%$ to 79.07$\%$) and lower variability (standard deviation of 1.544$\%$) compared to TRM-Ag (range of 73.07$\%$ to 80.28$\%$ and standard deviation of 1.956$\%$) (Table II). Addressing the contact resistance through further refinement of cavitation parameters or additional treatments, such as controlled annealing, could improve the fill factor.

Efficiency measurements showed comparable results for both pastes, with Cav-Ag achieving 21.43$\%$ and TRM-Ag achieving 21.58$\%$ (Table II). Despite the slightly lower efficiency, Cav-Ag exhibited more consistent performance with a narrower efficiency range (20.47$\%$to 21.88$\%$) and lower standard deviation (0.420$\%$) (Table II). Statistical analysis, including the Anderson-Darling test and T-test indcated no significant difference in mean efficiencies (p-value = 0.447) (Table \ref{tab:tab3}). This suggests that the consistent performance and potential cost savings make Cav-Ag a promising alternative for commercial production, provided the contact resistance could be further optimized.

\begin{table}[ht!]
\centering
\caption{Statistical Analysis of Cav-Ag and TRM-Ag Metallization Pastes}
\label{tab:tab3}
\begin{tabularx}{\columnwidth}{lXcccc}
\toprule
\textbf{Variable} & \textbf{Batch\_ID} & \textbf{N} & \textbf{P-Value} & \textbf{T-Test P-Value} \\
\midrule
Voc (V) & Cav-Ag & 12 & 0.236 & 0.000 \\
 & TRM-Ag & 12 & 0.217 &  \\
Isc (A) & Cav-Ag & 12 & 0.284 & 0.008 \\
 & TRM-Ag & 12 & 0.213 &  \\
FF (\%) & Cav-Ag & 12 & 0.057 & 0.034 \\
 & TRM-Ag & 12 & <0.005 &  \\
Efficiency (\%) & Cav-Ag & 12 & 0.168 & 0.447 \\
 & TRM-Ag & 12 & <0.005 &  \\
\bottomrule
\end{tabularx}
\end{table}

\begin{figure}[ht]
\centering
\includegraphics[width=.85\columnwidth]{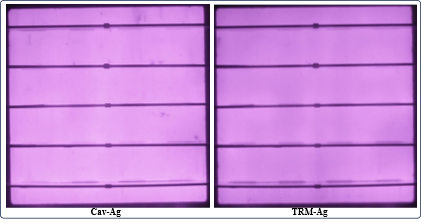}
\caption{Electroluminescence Analysis of Cav-Ag and TRM-Ag Paste}
\label{fig:el}
\end{figure}

Cav-Ag paste exhibited higher contact resistance (7.37E-02 $\Omega$-cm$^{2}$) compared to TRM-Ag (3.40E-03 $\Omega$-cm$^{2}$), and its series resistance was also higher (1.051 $\Omega$-cm$^{2}$ for Cav-Ag vs. 0.644 $\Omega$-cm$^{2}$ for TRM-Ag) as shown in Table \ref{tab:rcrs}. To address this, further optimization of the cavitation parameters, such as ultrasonic frequency and power, should be explored. Additionally, integrating addition post-procesing tehniques, beyond the initial sintering, like a controlled annealing process, could potentially reduce contact resistance. Electroluminescence (EL) imaging showed comparable uniform illumination for both pastes, indicating efficient charge carrier pathways (Figure \ref{fig:el}). The consistent performance of Cav-Ag, despite the higher contact resistsance, underscores the need to further research to enhance contact quality withouth compromising the benefits of fier gridlines.

\begin{table}[!h]
\centering
\caption{Series and Contact Resistance of Cav-Ag and TRM-Ag}
\label{tab:rcrs}
\begin{tabularx}{\columnwidth}{lXcccccc}
\toprule
\textbf{Batch\_ID} & \textbf{N} & {\textbf{Rs ($\Omega$-cm2)}} & {\textbf{Rc ($\Omega$-cm2)}} \\ 
\midrule
Cav-Ag & 12 & 1.051 & 7.37E-02 \\
TRM-Ag & 12 & 0.644 & 3.40E-03 \\ 
\bottomrule
\end{tabularx}%
\end{table}

This study underscores the potential of cavitation technology to enhance the production of silver paste for solar cells, reducing production costs and improving efficiency. The slight increase in V$_{OC}$ and I$_{SC}$, combined with consistent performance, makes Cav-Ag a promising alternative to TRM-Ag. Despite the challenge of higher contact resistance, further research and optimization could fully realize the benefits of this innovative approach, making it a viable option for future commercial applications in solar cell manufacturing.

%% file: Sections/Conclusion.tex
\section{Conclusion}

This study demonstrates that cavitation technology offers a promising and cost-effective alternative to traditional three-roll milling (TRM) in the production of silver paste for solar cells. The cavitation process significantly reduces particle size, enhances dispersion quality, and improves the overall properties of the paste. Cav-Ag paste showed finer gridlines, which contributed to a slight increase in open-circuit voltage (V$_{OC}$) and short circuit current (I$_{SC}$) due to reduced shading. While the fill factor (FF) was marginally lower for Cav-Ag, the paste exhibited more consistent performance and reduced variability across different metrics.

Despite the slightly higher contact resistance observed with Cav-Ag, the technology’s ability to produce uniform and finely dispersed particles presents a substantial advantage in solar cell manufacturing. The study underscores the potential of cavitation technology to lower production costs and improve efficiency, making it a viable alternative for future solar cell production. Further optimization of cavitation parameters and post-processing techniques are necessary to address the increased contact resistance. Future research should focus on refining these processes and exploring the integration of cavitation technology with other innovative methods to fully realize its benefits in commercial applications.

%% file: References/References.tex
\section{References}
\bibliography{References/aipsamp}